# Intensity-Based Registration of Freehand 3D Ultrasound and CT-scan Images of the Kidney

Antoine Leroy, Pierre Mozer, Yohan Payan, and Jocelyne Troccaz

*TIMC Lab – IN3S – Faculté de Médecine – Domaine de la Merci – 38706 La Tronche cedex – France*

+(33) 4 56 52 00 07

+(33) 4 56 52 00 55

antoine.leroy@koelis.com

pierre.mozer@psl.ap-hop-paris.fr

Yohan.Payan@imag.fr

Jocelyne.Troccaz@imag.fr

This work was partly supported by PRAXIM-Medivision, by the French National Agency for Research and Technology (ANRT) and by the non-profit organization "Association Française d'Urologie".

This paper presents a method to register a pre-operative Computed-Tomography (CT) volume to a sparse set of intra-operative Ultra-Sound (US) slices. In the context of percutaneous renal puncture, the aim is to transfer planning information to an intra-operative coordinate system. The spatial position of the US slices is measured by optically localizing a calibrated probe. Assuming the reproducibility of kidney motion during breathing, and no deformation of the organ, the method consists in optimizing a rigid 6 Degree Of Freedom (DOF) transform by evaluating at each step the similarity between the set of US images and the CT volume. The correlation between CT and US images being naturally rather poor, the images have been preprocessed in order to increase their similarity. Among the similarity measures formerly studied in the context of medical image registration, Correlation Ratio (CR) turned out to be one of the most accurate and appropriate, particularly with the chosen non-derivative minimization scheme, namely Powell-Brent's. The resulting matching transforms are compared to a standard rigid surface registration involving segmentation, regarding both accuracy and repeatability. The obtained results are presented and discussed.

*Computer Assisted Surgery, Kidney, Intensity-based registration, CT-scanner, Freehand Ultrasound.*

# INTRODUCTION

**Clinical and Scientific Context**

Percutaneous Renal Puncture (PRP) is a common surgical procedure that precedes the extraction of renal stones through a nephroscope. It consists in percutaneously inserting a long and thin needle into one of the narrow calices of the kidney under intra-operative fluoroscopic or ultrasound control. This procedure is an alternative to lithotripsy or ureteroscopy when the stones are too numerous, too big or hardly reachable.

The pre-operative imaging modality is CT. An early acquisition of the kidneys after injection of a contrast product allows seeing clearly the kidney parenchyma, whereas a late acquisition enhances the tree-like internal cavities (see figure 1). Intra-operatively, either fluoroscopy or ultrasound imaging is used for target visualization. So the surgeon has to perform a mental matching between the static horizontal slices from CT and the dynamic intra-operative images to find the targeted cavity, and to decide which way is best to reach it percutaneously. The interviewed surgeons believe that PRP accuracy could benefit from computer assistance, to cope with the lack of visibility, the poor quality of ultrasound images, the irradiating nature of X-Rays and with the difficulty to mentally represent the 3D localization of the small target inside the kidney from those imaging modalities.

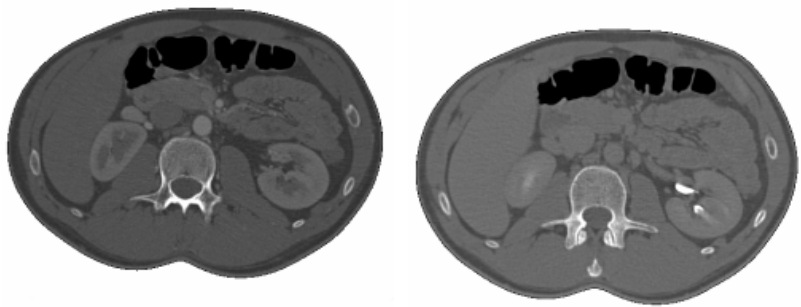

**Figure 1: CT early (left) and late (right) acquisitions**

Therefore, a feasibility study on Computer-Assisted PRP was previously carried out [1,2], in which the kidney surface, segmented from CT and localized US images, was registered using a surface registration algorithm, enabling navigation of the needle. The study showed satisfying results briefly reported in section "Overview of the Method" below; however it required the intra-operative segmentation of US images, which automation is very difficult and for which an interaction with the surgeon is not practicable for a clinical use.

Therefore, automated CT/US registration has been investigated. It was decided for the present study to propose and evaluate a voxel-based registration algorithm, in order to avoid segmentation steps and to minimize user intervention. The developed method has been compared to the surface-based registration of the same data.

**State of the Art**

Voxel-based registration methods have been thoroughly studied since 1995. Every method proposes a similarity measure and a cost minimization algorithm. [3] first introduced Mutual Information (MI) combined with histogram windowing and a gradient descent algorithm. [4] presented an interesting combination of MI and Powell-Brent (PB) search strategy. [5] compared various searches and multi-resolution strategies, and showed that PB was efficient with image subsampling. [6], [7] and [8] made thorough comparisons of different functional and statistical similarity measures. [9] and more recently [10] proposed to add gradient information for similarity computation.

Although those studies constitute the base of our research, none of them specifically involves US images. We therefore focused on the works of Roche et al. [11,12], who registered 3D US of the brain with MRI, and Penney et al. [13], who registered 2.5D freehand US of the liver with MRI. The theoretical difficulty in registering CT and US is that the former gives information on tissues intensity, whereas the latter contains a speckled image of their boundaries. So a complex similarity measure and specific image preprocessing must be chosen.

In that context, Penney and colleagues' work [13] is based on probability maps: those maps associated to the data to be registered correspond to the probability of each pixel/voxel to belong to the hepatic tree or to the parenchyma; those transformed data built from a learning phase being mono-modal can thus be registered using a simple normalized cross-correlation. This approach is a very elegant trade-off between contour-based registration requiring the accurate segmentation of data and intensity-based ones that measure image mutual dependency. It benefits from the fact that the hepatic tree is quite visible and different from the parenchyma. Such an approach is not easily applicable to kidney images. Our choice was therefore to process CT and US images before registration to make them more similar whilst keeping in the multi-modality framework. However the method we propose shows several similarities to the one presented by Penney and colleagues: rigidity of the organ assumed, breathing cycle taken into account, specific preprocessing of images including acoustic shadow removal, non-derivative iterative optimization scheme, manual point-to-point pre-registration.

**Overview of the Method**

This paper introduces a method to automatically rigidly register localized US images of the kidney onto a high-quality abdominal CT volume. This method is intended to be integrated in a navigation system (see [1,2]) where pre-operative planning is performed on two registered CT scans; US data are acquired intra-operatively during patient apnea, and then registered to CT data; finally navigation of the puncture needle is performed in the same breathing state than US data acquisition. The US probe and needle are localized using a passive optical device (Polaris® from NDI).

The selected intensity-based registration method uses image preprocessing in both modalities, Powell-Brent method as a search strategy, and Correlation Ratio (CR) as a similarity measure. Tri-linear interpolation is used when superimposing one data set to the other before CR computation.

Preliminary tests have been carried out and are compared to a "Bronze Standard[1]" transform (BS) - as defined in [14] based on 3D/3D surface matching [1,2] (fig. 2) using octree distance maps and Levenberg-Marquardt minimization.

---

[1] By "Bronze standard' we mean a method thoroughly evaluated and that can be considered as a reference but which is not supposed to be the ground truth as is a "gold standard" (based for instance on the use of artificial or anatomical landmarks).

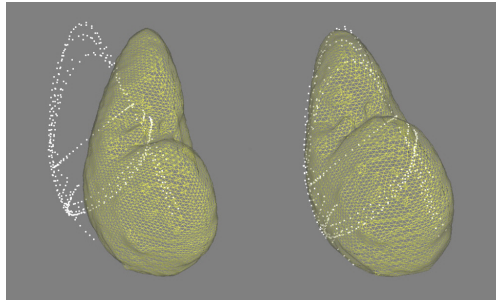

**Figure 2: Rigid surface registration of US and CT segmented kidney (left: before registration; right: after registration).**

The assessment of surface-based registration on real data was performed in different ways: by testing reproducibility from different initial attitudes (initial errors up to +/-30mm, +/-30°); by testing coherency in a registration circuit (early-injection CT to US, late-injection CT to US, early-injection CT to late-injection CT); and by computing residual distance between registered data sets on surface, centroid and axis of the kidney. For initial attitudes having their 6 parameters smaller or equal to 20 (mm or °), the convergence and coherency are very good and the average residual distances are respectively 1mm (surface), 1.4mm (centroid) and 3.6° (axis). Including initial errors up to 30 (mm and °) gives in average 1.6mm, 3.8mm and 8.4°.

Additional tests on a phantom including needle guidance and comparison from post-operative CT data between planned trajectories and inserted needles gave in average an error of 4.7mm measured at the needle tip; this error cumulates the system error and the additional errors due to the segmentation and registration processing steps of post-operative CT data. The analysis of those experiments showed that a major part of the global error is due to needle flexion during insertion in the phantom (let us remind that the use of the optical localizer makes the direct tracking of the needle tip impossible).

This surface registration serves as a standard for comparison to the method presented in the paper.

## REPRODUCIBILITY OF KIDNEY MOVEMENTS

The approach assumes that the kidney is in the same position when getting intra-operative US data and for percutaneous puncture; both being performed during apneas of the patient. As the kidney moves with respiration, the problem is therefore to verify how much this motion is reproducible from one breathing cycle to another.

Langen and Jones [15] report several clinical studies of kidney motion using different modalities: CT, US, MRI, X-rays. All those studies agree on the fact that the motions amplitude of the kidney is in average between one and two centimeters during normal breathing but no information is given about the repeatability of those motions. Schwarz and colleagues [16] report a study on 14 patients from fast MRI data, the measurements being made on two acquisition planes: in average the kidney motion is about 17mm and its reproducibility is 3mm. However, we believe that measurements on two low-resolution MRI slices cannot give accurate values on the 3D motion of the kidney.

To complement this knowledge and verify the assumption concerning the reproducibility of kidney movements, experimental work was carried out using localized US images. Static and dynamic acquisitions were made on 11 volunteers. The static acquisitions consisted in sweeping the localized US probe over the kidney whilst each volunteer was carrying out 3 experiments of full inhales (50 US images recorded per experiment) and 2 experiments of full exhales (30 US images per experiment). For each experiment, US images were manually segmented leading to a 3D set of data points located at the surface of the kidney. By registering two sets of surface points coming from two different experiments using the rigid surface registration algorithm described in the previous section, it was possible to estimate the kidney displacement between inhale and exhale (one inhale experiment compared to one exhale experiment) as well as the reproducibility of the kidney position for each breathing cycle (one inhale — resp. exhale — experiment compared to another one inhale — resp. exhale — experiment).

To complete the estimation of the kidney displacement between inhale and exhale provided by these static acquisitions, dynamic acquisitions were carried out consisting in aligning the ultrasound plane to the main plane of kidney motion so as to see the longitudinal axis of the kidney, and in acquiring its motion during five breathing cycles (leading to the record and segmentation of 100 dynamic US images per subject).

To make the kidney motion easier to visualize and to interpret, the displacements of two points – the centroid G and the pole P – as well as the inclination α of the principal axis of the organ were measured from those acquisitions (fig. 3).

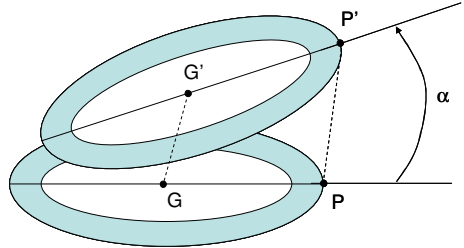

**Figure 3: Kidney motion parameters. GG' and PP' are the displacements of respectively the centroid (G) and the pole (P) of the organ; α is the inclination from exhale to inhale position.**

Table 1 summarizes the results obtained for the estimations of the kidney displacement during one breathing cycle.

Concerning the evaluation of the reproducibility of the kidney position for each inhale or exhale (Table 2), it was decided to focus on GG' distance as α is rather small and PP' values are rather similar to GG' ones.

**Table 1. Measured amplitude of the kidney motion for 11 subjects (in mm and degrees).**

|     | mean | Std.dev | min  | max  |
| --- | ---- | ------- | ---- | ---- |
| PP' | 30.8 | 14.1    | 7.8  | 55.5 |
| GG' | 30.1 | 14.7    | 10.1 | 60.0 |
| α   | 11.6 | 3.8     | 6.5  | 17.5 |

**Table 2. Reproducibility of centroid position for each breathing cycle (in mm).**

|     | mean | Std.dev | min | max  |
| --- | ---- | ------- | --- | ---- |
| ΔG  | 5.9  | 3.9     | 1.3 | 15.9 |

In terms of position repeatability the worst results were obtained for two subjects having the largest index of body mass and for which the ultrasound acquisition and processing turned out to be quite difficult. The best results were obtained for a sportsman having very good control of his apneas, with a measured average reproducibility of 1.8mm. Those results are coherent with published studies and encouraging concerning our hypothesis that the kidney motion is reproducible. In operative conditions, using a breathing machine, the results should be improved. A study on patients is envisioned, as well as the use of localized 3D ultrasound.

# MATERIAL

## CT images

CT-scan images are organized in a stack of parallel horizontal slices. Every voxel of the CT volume conveys information on the X-ray intensity at a given position in space. So, homogeneous regions or organs in the body appear as homogeneous grey-level regions in the slices. The volumes we worked on have a typical size of 512x512x300 voxels, with an inter-slice space of 0.625mm and a slice thickness of 1.25mm. Thanks to dedicated software (Analyze™), the data were first re-sampled so that the voxel be cubic, with a 0.625mm width.

As was already mentioned, more specifically to the application, two different CT scans are available: early- and late-acquisition scans. In the early scan, the contrast product fills the highly-vascularized kidney parenchyma; the kidney external shape is thus more easily segmentable, thanks to a homogeneous higher Hounsfield value. In the late scan, after a few minutes, the contrast product flows into the internal kidney cavities, hence providing an easy delineation in the images of a bright tree-like structure inside the organ (see figure 4).

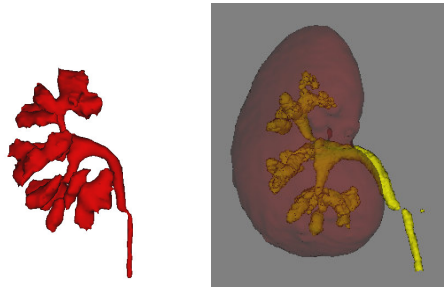

**Figure 4: Right kidney with its internal cavities. (Left) internal cavities segmented from late CT acquisition and (right) repositioned with respect to the kidney external surface segmented from the early CT acquisition.**

The kidneys are first semi-automatically segmented using Nabla's 3D watershed (from Generic Vision Inc.) in the two CT scans; then internal cavities may be extracted for planning from the late scan by thresholding. The registration of the kidney from early to late acquisition allows the surgeon to perform a 3D diagnosis and pre-operative planning, with both the information of the kidney external shape (used for computation of the reference registration) and cavities (where the target is defined). This mono-modal registration is performed using a classical chamfer rigid matching based on the kidney surface visible on both CT acquisitions.

In the current study, the late-acquisition CT scan is selected for registration to intra-operative data since it showed a better correlation with the US images than the early-acquisition CT.

**US images**

US images convey information on the acoustic interfaces between the organs. They are sometimes called "gradient images" for that reason. The acoustic waves that propagate through the soft tissues emit a reflected echo at each interface and lose intensity with increasing distance to the probe. On the contrary, structures with a higher reflection coefficient (bone, air, renal stone, etc.) tend to reflect the totality of the signal, thus creating an "acoustic shadow" behind them w.r.t. the probe. US images are known to be highly noisy, the noise called "speckle" being generated by the numerous micro-reflections over the swept area.

Not only the interfaces between regions are represented in US images, but also the regions themselves, to a certain extent, since every region has its own internal cell organization that generates a specific speckle signal. This property should make the comparison between CT and US images feasible.

The US images are calibrated and localized in space with the help of an optical localizer. Calibration is performed using Lango's method [17]. The acquisition of dozens of localized images allows building a sparse set of US voxels in space, that we aim at registering with the dense volume of CT data. The images acquired through an analog video board have a typical size of 768x576 pixels, the pixel being 0.3x0.3mm (US device: Hitachi-EUB-405, 3.5 and 7.5MHz probes).

# METHOD

Intensity-based registration is an iterative method to search for an optimal transform T between two sets of data V1 and V2 – in this case $V_{CT}$ and $V_{US}$; the optimality being defined by a similarity measure between the pixels/voxels of the two data sets.

As previously introduced, the method starts with a preprocessing of the images to make them more similar. Due to the presence of local minima in the optimization process, pre-registration is necessary for robust initialization of T; the initial value of T, called initial attitude (IA) is computed by a landmark registration. Because the environment of the kidney may be deformed (liver especially), a rigid registration on the whole image would be inappropriate. Therefore, the similarity is computed in a special region of interest (ROI), delineated by the user "around" the kidney. The ROI is defined as the smallest parallelepiped parallel to the US image axes that includes the kidney and its boundaries on the image. As regards the registration process itself, a similarity measure among those available, and an adequate optimization scheme must be chosen.

This may be summarized by the following equation:
$$T_{opt} = \arg\max_{T} (similarity_{ROI} (V_{CT}, T(V_{US})))$$
where $V_{CT}$ is the CT data volume and $V_{US}=\{S_1, S_2, …, S_k\}$ the set of localized US slices. Let us remark that US data are mapped to CT ones since the CT volume is dense whilst US slices are sparse data.

**Preprocessing**

*CT-scan images*

The goal of this stage is to highlight the major boundaries in CT in order to increase the similarity with the "gradient" US images.

Penney and colleagues [13] basically transformed the blurred MRI and US images of the liver into maps giving the probability of a pixel to be liver tissue or vessel lumen. However, as this process requires a manual thresholding of the MRI, and as the segmentation of the kidney parenchyma is not reduced to a binarization process, especially in the US images, this technique cannot apply to the kidney problem. Roche et al. [12] proposed the combined use of the MRI image and its derivative. Again, it was decided not to use this approach because of the complexity of the bivariate correlation method and because of the observed chaotic correlation between CT gradient and US.

The selected approach is the following: the preprocessing of a CT slice consists in the superimposition of a median blur and a bi-directional Sobel gradient filter. After contrasting and thresholding the gradient image, the largest connected components are kept and superimposed to the blurred image (fig. 5). Sobel filter is adequate here since it creates fairly wide boundaries, whose contribution in the similarity measure will lead to deeper minima in the cost function.

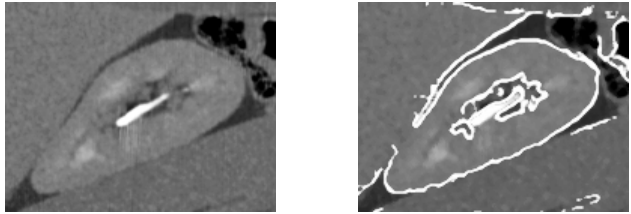

**Figure 5: Oblique CT kidney after preprocessing: low-pass and contour delineation. (Left) before; (right) after.**

*US Images*

*Speckle Removal.* US images are known to be blurred by speckle noise. Still, the kidney, due to its small size and echogenic capsule, can be well and fully visualized through anterior access. The aim of preprocessing US images is to reduce the speckle, while preserving the echogenic boundaries of the organ.

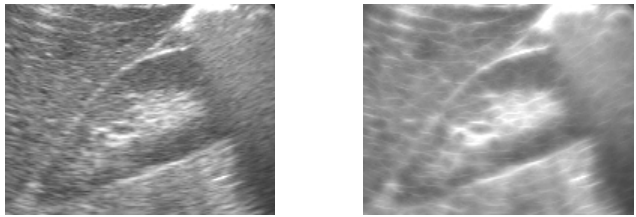

**Figure 6: Preprocessing of a US slice using sticks. Most speckle is removed while boundaries are preserved. (Left) before; (right) after.**

The Sticks filter, which use for speckle removal was theoretically justified by Czerwinski et al. [18], was experimentally compared as regards registration to two other filters taken from the literature (median and "salt and pepper" Crimmins'). We finally chose the Sticks filter since, while performing a fairly efficient speckle removal, it kept the bright and "high frequency" boundary information of the kidney and this resulted in improved registration convergence. This filter consists, over an "n x n" convolution window, in computing the maximum of the average grey level over each oblique stick in the window. An example result is shown on figure 6.

*Shadow Removal.* Because they convey no relevant information for intensity-based registration, acoustic shadows are removed in the US images, as proposed in [13]. The shadows are produced by large interfaces like the ribs or the colon that quasi-totally reflect the acoustic signal, the remaining US waves decreasing in the distance in an exponential way. In other words, the shadow generally follows a maximum of the intensity profile along a US scan-line. Shadow removal is therefore based on the correlation between a US scan-line profile and a heuristic exponential function. Figure 7 shows grey level values along a scan line. The definition of the heuristic function can be found in the appendix.

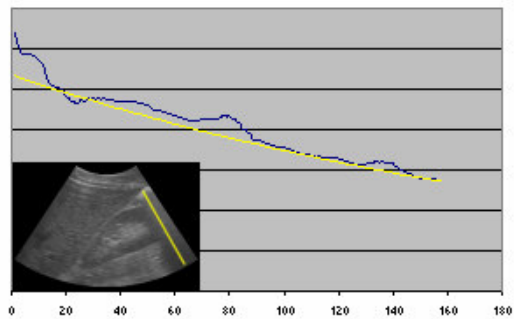

**Figure 7: Grey profile along the US scan line profile (dark) compared to the heuristic correlated function (pale); the corresponding direction indicated in the US image includes a shadow. X is the pixel position along the line – Y is the pixel intensity.**

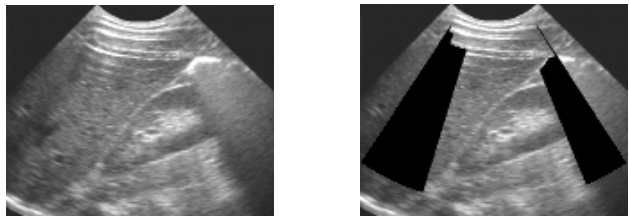

**Figure 8: US shadow removal. Left image shows the shadow profiles induced by the ribs (left) and the colon (right). The generated mask is superimposed on right image.**

A shadow is detected when the correlation coefficient is higher than a given threshold, and when the maximum acoustic interface is within a threshold distance to the skin. Figure 8 shows the result of this automatic method on a sample slice. Those thresholds have been selected from experiments on series of typical US images.

*Effect of preprocessing on registration*

Preprocessing effect upon registration was tested on a set of US slices and their corresponding CT oblique slice. Figure 9 and 10 present a typical example.

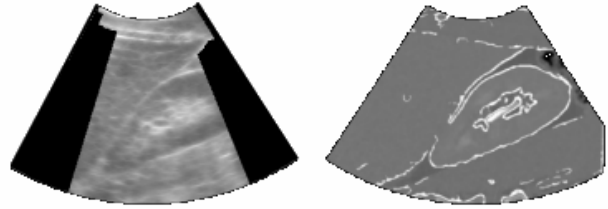

**Figure 9: A sample US slice and its corresponding CT oblique slice, after preprocessing.**

A systematic 2D registration test with an exhaustive exploration of the searching space showed that the cost function for the correlation ratio presents an obvious global minimum when the images are preprocessed (fig. 10).

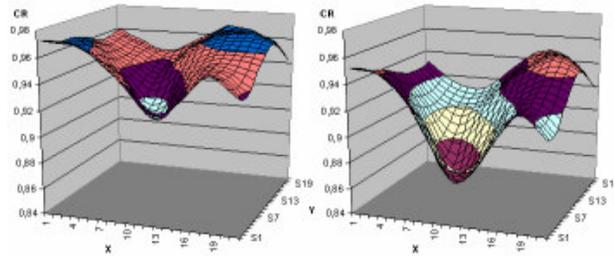

**Figure 10: Sample 2D registration of corresponding CT/US slices using CR, without preprocessing (left) and with preprocessing (right). The attraction basin is deeper in the latter case.**

## Similarity Measure: Correlation Ratio

*Definition*

Let X be the base image (CT) and Y the floating image (US) to be matched to X. The CR is a functional similarity measure [6,8,11] based on conditional moments. It measures the functional relationship between the floating pixels and the base ones. Roche and colleagues [11] define the CR as:

$$\eta(Y|X) = 1 - \frac{1}{N\sigma^2}\sum_a N_a \sigma_a^2$$

where N is the number of overlapping pixels of X and Y for a given value of the unknown transform T, $\sigma$ their variance in Y; $N_a$ is the number of overlapping pixels in X which grey intensity value is a, $\sigma_a$ their variance in Y. A more complete definition of the CR is given in the appendix.

*Motivation*

Theoretically, the functional relationship between an intensity CT image and a gradient US image is not obvious, thus a statistical similarity measure like normalized mutual information (NMI) should be more appropriate. However, although US images best enhance the boundaries between organs, they also convey information on the tissues, as each of them diffuses the US waves its own way, in both intensity and frequency spaces. It was therefore assumed that CR is somehow justified, especially with the preprocessing steps presented above. We also chose CR against NMI for several reasons:

- CR looks smoother than NMI (fig. 11); few local minima appear, even when sub-sampling: it is more accurate, discriminative and robust [8].
- Both CR and NMI algorithm complexities are O(N), where N is the number of overlapping pixels in the ROI; however CR requires less iterations, and is thus 30% to 60% faster.
- CR smooth profile is particularly adapted to PB search algorithm, since Brent minimization is based on the fitting of a parabola onto the function curve, as is described in the "Optimization scheme" section [19].

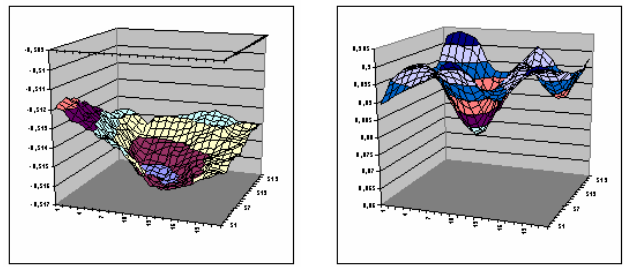

**Figure 11: NMI (left) and CR (right) profiles mapped along $T_X$ and $T_Y$ while registering 2 preprocessed images. Note the smooth aspect of CR and the deep unambiguous minimum.**

*2.5D Correlation Ratio*

In the context of 2.5D US, the CR is computed as follows: for every rigid transform estimate, the CT volume is re-sliced in the plane of each US slice, using tri-linear interpolation. Then the region of interest (ROI) defined in each US image is superimposed onto the corresponding CT oblique slice. Finally, the CR is globally estimated on the set of pixels from the superimposed ROIs. In other words, the different US images are considered as a whole sparse set of pixels.

**Optimization Scheme: Powell-Brent**

PB is an iterative multi-directional search, based on the iterative fitting of a parabola to each 1D profile of the cost function in order to find 1D minima [4,6]. It appears as a fairly efficient search strategy when the differentiation of the cost function is unknown [4]. Our implementation is based on [19]. Some changes were applied since it was found that the method was too slow and too sensitive to local minima, which are frequent in image registration.

*Initial Attitude and Parameter Space*

The minimization process consists in optimizing a 6D position and orientation vector. The initial attitude (IA) is defined by a point-to-point registration using the Arun algorithm [20]. Those points are anatomical landmarks visible on the US and CT data. The two poles (extreme points of the kidney along the CT Z axis) are identified and chosen on the CT images and on the US images. The other landmarks (at least one more) are generally chosen from a median transverse slice of the organ, for example at the connection with the ureter, or on the other side on the back of the kidney. On our data set the three or four points cited above where relatively well identifiable and repeatable. A reference system attached to the kidney is defined from those points: for instance the longitudinal axis of the CT kidney is defined as the Z axis of the parameter space. Finally, as Maes and colleagues suggests in [5], the better-conditioned parameter vector $(T_X,T_Y,R_Z,T_Z,R_X,R_Y)$ is used for optimization.

*Search Interval limiting*

In the original PB method, Brent 1D search interval for each DOF is defined by a time-consuming automatic bracketing. Based on repeated experiments about landmark selection for initial pre-registration and convergence from computed initial attitudes, an initial interval length of $2*RMS_{ARUN}$ (20 to 30mm) around the IA was chosen.

*1D Initial search*

Jenkinson et al. [6] proposed an initial systematic search on the rotational DOFs, previously to PB iterations, to make the solution "more reliable". In the same idea, a two stage approach was adopted with two step values. As PB method tends to converge to local minima, and as the similarity between registered CT and US shows in most cases a non-ambiguous global minimum (fig. 11), we chose to perform an initial search before each Brent 1D optimization, with a step of 10% of the interval length, to find an approximate global minimum in that DOF (fig. 12 left). Then Brent method is applied around that minimum in a very restricted interval (fig. 12 right). With this two stage approach, we observed that local minima were considerably avoided.

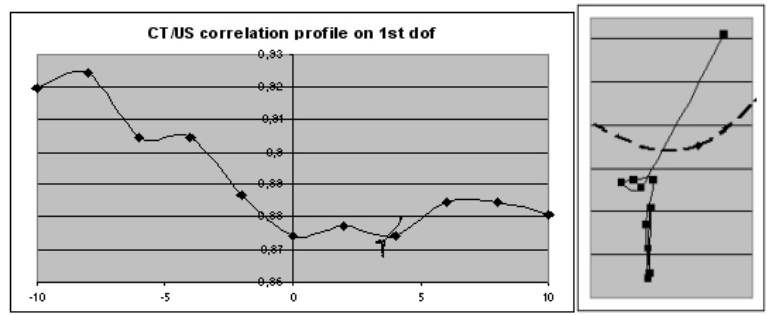

**Figure 12: CR profile along the search interval for $T_X$. The converging curve represents Brent iterations around the global minimum of the profile.**

*Search interval reduction*

Since PB tends to reach the solution very fast (only the first iteration makes a significant step), it was decided, not only for computation time improvement but also to avoid peripheral local minima, to reduce the interval length at each Powell iteration with a factor of 1.5, until a certain limit is reached.

*Normalizing search directions*

After each Powell iteration, the new conjugate direction is normalized instead of being multiplied by the norm of the last found 6D minimum; indeed this minimum can be very small and can lead to early convergence in a local minimum.

# RESULTS

**Materials and Conditions**

Several evaluations were performed on a single data set. This specific isotropic abdominal CT volume has 517x517x287 voxels, the voxel size being 0.625mm. Our registration experiments were carried out on a late acquisition CT, after iodine injection. Besides, for those experiments, 100 US images (axial and longitudinal) of the right kidney were acquired with a localized calibrated US probe, through anterior access. Those images which pixel size is 0.3mm are then resized to match CT voxel size in order to make further processing faster. In the final protocol, in order to stay in a framework compatible with a clinical use, we chose to carry out each registration on only 5 slices of various orientations out of the whole set. The ROI – parallelepipeds – are chosen to be tangent to the US kidney.

**Measured Data**

The accuracy study consisted in performing two series of 10 registrations: on the one hand, by fixing the IA and changing the set of 5 US slices to be matched to the CT scan; on the other hand, by computing a new IA at each attempt, while keeping the same set of slices. Every IA is computed from the 4 anatomical points determined by the user on CT and US data at the axial and lateral extremities of the kidney.

Let us remind that a reference transform $T_{BS}$ has been computed from the surface registration (cf. "Overview of the Method" section). The quality of a computed transform T is characterized by three measures (cf. fig. 3):

- The distance between the two CT meshes obtained by transforming the CT data respectively by $T_{BS}$ and T. This distance is computed between corresponding points of the initial mesh; mean and standard deviation values are presented for each registration trial.
- The distance between their centroids.
- The difference between their principal axes (angular distance).

Considering those three measures gives a better appreciation of the registration accuracy and success than a single RMS on registered data sets (CT and US).

**Registration results**

The registrations took on average 80s[2] (from 11s to 170s). The algorithm appeared to be stable since a second registration (i.e. another registration started from an IA that is the attitude provided by the first registration) kept the kidney with the same attitude.

**Table 3 Matching statistics for varying set of US slices and constant IA (in mm and degrees).**

| Trial nb | Meshes distances (mean+/-std dev) | Centroid distances | Axis angular distances |
|---|---|---|---|
| 1 | 5.6 (+/- 2.1) | 2.7 | 4.1 |
| 2 | 6.8 (+/- 2.0) | 3.2 | 6.7 |
| 3 | 5.3 (+/- 1.5) | 3.2 | 1.3 |
| 4 | 6.2 (+/- 1.8) | 4.0 | 5.3 |
| 5 | 5.8 (+/- 1.8) | 2.3 | 9.5 |
| 6 | 6.0 (+/- 2.0) | 4.3 | 4.7 |
| 7 | 4.1 (+/- 1.1) | 2.7 | 5.5 |
| 8 | 5.1 (+/- 1.1) | 3.4 | 5.2 |
| 9 | 5.1 (+/- 1.9) | 3.0 | 5.4 |
| 10 | 3.7 (+/- 0.9) | 2.1 | 2.9 |
| Average | 5.36 +/- 0.93 | 3.09 +/- 0.69 | 5.06 +/- 2.17 |

**Table 4 Matching statistics for a constant set of US slices and varying IA (in mm and degrees).**

| Trial nb | Meshes distances (mean+/-std dev) | Centroid distances | Axis angular distances |
|---|---|---|---|
| 11 | 3.7 (+/- 1.3) | 1.3 | 4.3 |
| 12 | 3.2 (+/- 0.8) | 2.9 | 1.5 |
| 13 | 8.8 (+/- 2.5) | 5.7 | 10.5 |
| 14 | 5.7 (+/- 2.0) | 4.0 | 5.3 |
| 15 | 6.7 (+/- 2.6) | 5.1 | 8.2 |
| 16 | 6.0 (+/- 1.3) | 5.8 | 3.2 |
| 17 | 11.9 (+/- 4.8) | 8.1 | 6.4 |
| 18 | 6.5 (+/- 2.1) | 5.0 | 7.6 |
| 19 | 6.2 (+/- 1.6) | 5.6 | 6.0 |
| 20 | 5.4 (+/- 0.9) | 4.1 | 4.3 |
| Average | 6.41 +/- 2.47 | 4.76 +/- 1.83 | 5.73 +/- 2.61 |

---

[2] Tests performed on a Pentium IV 1.7GHz

Tables 3 and 4 show matching statistics using the three measures listed above. On average, the IA computed from point-to-point registration sets the kidney at 7.7mm±3.5 in translation and 9.2°±4.0 in inclination from the reference solution. With 1mm to 6mm error in translation, and 1° to 7° in inclination, 15 registrations out of 20 have been considered successful. Figure 13 shows one of them. Trials 13 and 17 are considered as failures, whilst trials 5, 15 and 18 present higher errors on the kidney inclination than the average.

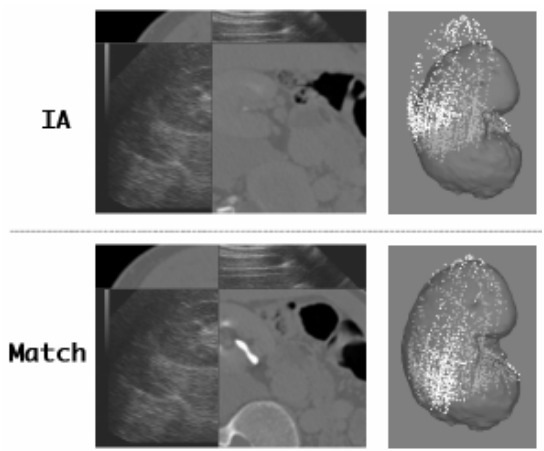

Figure 13: Example of successful registration, showing superimposed CT/US slices and meshes. (Top): after determining the Initial Attitude using anatomical landmarks – (bottom) after registration using the intensity-based method.

## DISCUSSION

We presented an original CT/US voxel-based registration method, dedicated to kidney surgery. The general approach, specific choices and results are discussed in this section.

The preprocessing of medical images before registration is sometimes put into question, since the cost function does have a global minimum. By preprocessing data to be registered we chose to deepen the attraction basin of that minimum. The side effect is that local minima also get deeper, even in a 20mm-radius neighborhood. However, by performing a systematic pre-search of the global minimum along every DOF, the risk of misregistration is decreased.

At any case, the removal of acoustic shadows appeared to be efficient in reducing the number of local minima. The method that was presented is fast and robust, provided that the quality of the images is good enough, in terms of contrast and regions homogeneity, particularly. This is an intrinsic drawback of using US data for which the image quality may be very variable from patient to patient but this drawback is the same when performing the puncture in the traditional way. Robustness to patient variability could probably be improved by integrating hybrid methods alternating registration and automatic segmentation stages such as presented in [21,22].

As regards the similarity measure, the CR seemed to be the best choice to make in the context of this application. Aside from the computation time question, the NMI as well as the other "statistic" and not "functional" measures presented a profile not sufficiently smooth to be used. Although the systematic search may bring closer to the large attraction basin of the global minimum, in the final iterations Brent's algorithm is very likely to converge to one of the numerous surrounding local minima. Powell-Brent optimization scheme is much more successful with a smoother cost function such as CR.

Regarding the minimization process, a non-derivative method was chosen. Two well-known methods were good candidates: Powell-Brent and the Nelder-Mead simplex. The former was selected as it allows a better control of the convergence, of the search domain, etc. With a correct implementation, the computation time is not greater for PB. However, authors like Maes et al. [5] have been able to use mutual information in derivative processes; some ideas may be taken out of it.

The developments were made, as well as several types of experiments, on one single data set from a young man; US data have been acquired through anterior access, which provides clear images. Some implementation choices were specific to this anterior access, particularly as regards the preprocessing of the images. Using the more traditional posterior access could make necessary the choice of some other preprocessing tools; however the general approach can be kept identical.

Concerning accuracy, an average distance of 5.36mm to a bronze standard registration has been obtained where Penney and colleagues obtain in average 3.6mm. Whilst not totally satisfactory, those measurements which constrain the size of the target to be reached are reasonable in the context of this application (access to the pyelocaliceal cavities which size averages 1cm). This accuracy has to be confirmed on larger sets of data before envisioning the next stage involving needle guidance.

## CONCLUSION

In this paper we addressed the problem of registering a CT-scan volume with a sparse set of localized US slices of the kidney. The intensity-based registration method is threefold: a specific preprocessing of the images, a suited similarity measure and an efficient optimization process have been implemented. The preprocessing consists in decreasing the noise, enhancing the boundaries, and removing acoustic shadows. As for the similarity measure, correlation ratio was chosen against mutual information for its continuity and its computation time. The non-derivative Powell-Brent algorithm was implemented and allowed a precise control of the minimization process. Experimental results were compared to a "bronze standard" solution. Because of the theoretical difficulties in matching CT with US images, the presented results are satisfying in regard with accuracy and computation time. Nevertheless, we must bear in mind that their performances still depend on the choice of slices, the manual ROI, the manual IA and the search interval. Improving the robustness of the method with respect to those stages is a very important issue to focus on in the future.

## Appendix

**Heuristic function used for shadow removal**

Let $E = \{X_1, X_2, \ldots, X_k, \ldots, X_N\}$ be the profile of a scan-line in the US image. $X_k$ is a grey level value along the profile. If i is the index of maximum grey level, let $\mu$ and $\sigma$ be the mean and standard deviation of $E' = \{X_i, X_{i+1}, \ldots, X_N\}$. For k in [i,N], let define $F = \{Y_i, Y_{i+1}, \ldots, Y_N\}$ where $Y_k = f(k)$. The heuristic exponential function $f$ is defined as:

$$f(k) = A\exp(ak); A = \mu + \frac{3}{2}\sigma; a = \frac{1}{N}\ln(X_N/A)$$

The correlation coefficient is:

$$\Gamma = \frac{\left|\frac{1}{N'}\sum_{k=i}^{N} X_k Y_k - m\right|}{\sqrt{\left|\left(\frac{1}{N'}\sum_{k=i}^{N} X_k^2 - m\right)\left(\frac{1}{N'}\sum_{k=i}^{N} Y_k^2 - m\right)\right|}}, m = \left(\frac{1}{N'}\sum_{k=i}^{N} X_k\right)\left(\frac{1}{N'}\sum_{k=i}^{N} Y_k\right)$$

**Definition of the correlation ratio**

Let X and Y be two sets of pixels/voxels. Ω is the overlapping set of pixels/voxels and N its cardinal. For a grey level a present in X, we note $\Omega_a = \{i \mid X(i) = a\}$ the set of pixel/voxel indexes in X having this grey level intensity and $N_a = \text{card}(\Omega_a)$

Let $\sigma^2 = \frac{1}{N}\sum_{j \in \Omega} Y(j)^2 - m^2$ and $m^2 = \sum_{j \in \Omega} Y(j)$

$\sigma_a^2 = \frac{1}{N_a}\sum_{j \in \Omega_a} Y(j)^2 - m_a^2$ and $m_a^2 = \sum_{j \in \Omega_a} Y(j)$

The correlation ratio is $\eta(Y|X) = 1 - \frac{1}{N\sigma^2}\sum_a N_a \sigma_a^2$

FIGURE CAPTIONS

Figure 1: CT early (left) and late (right) acquisitions

Figure 2: Rigid surface registration of US and CT segmented kidney (left: before registration; right: after registration).

Figure 3: Kidney motion parameters. GG' and PP' are the displacements of respectively the centroid (G) and the pole (P) of the organ; $\alpha$ is the inclination from exhale to inhale position.

Figure 4: Right kidney with its internal cavities. (Left) internal cavities segmented from late CT acquisition and (right) repositioned with respect to the kidney external surface segmented from the early CT acquisition.

Figure 5: Oblique CT kidney after preprocessing: low-pass and contour delineation. (Left) before; (right) after)

Figure 6: Preprocessing of a US slice using sticks. Most speckle is removed while boundaries are preserved. (Left) before; (right) after)

Figure 7: Grey profile along the US scan line profile (dark) compared to the heuristic correlated function (pale); the corresponding direction indicated in the US image includes a shadow. X is the pixel position along the line – Y is the pixel intensity.

Figure 8: US shadow removal. Left image shows the shadow profiles induced by the ribs (left) and the colon (right). The generated mask is superimposed on right image.

Figure 9: A sample US slice and its corresponding CT oblique slice, after preprocessing.

Figure 10: Sample 2D registration of corresponding CT/US slices using CR, without preprocessing (left) and with preprocessing (right). The attraction basin is deeper in the latter case.

Figure 11: NMI (left) and CR (right) profiles mapped along $T_X$ and $T_Y$ while registering 2 preprocessed images. Note the smooth aspect of CR and the deep unambiguous minimum.

Figure 12: CR profile along the search interval for $T_X$. The converging curve represents Brent iterations around the global minimum of the profile.

Figure 13: Example of successful registration, showing superimposed CT/US slices and meshes. (Top): after determining the Initial Attitude using anatomical landmarks – (bottom) after registration using the intensity-based method.